\documentclass[12pt,letterpaper]{article}

\usepackage{fixltx2e}
\usepackage{textcomp}
\usepackage{fullpage}
\usepackage{amsfonts}
\usepackage{verbatim}
\usepackage[english]{babel}
\usepackage{pifont}
\usepackage{color}
\usepackage{setspace}
\usepackage{lscape}
\usepackage{indentfirst}
\usepackage[normalem]{ulem}
\usepackage{booktabs}
\usepackage{placeins}
\usepackage{natbib}
\usepackage{float}
\usepackage{latexsym}
\usepackage{url}
\usepackage{hyperref}
\usepackage{epsfig}
\usepackage{graphicx}
\usepackage{amssymb}
\usepackage{amsmath}
\usepackage{bm}
\usepackage{array}
\usepackage[version=3]{mhchem}
\usepackage{ifthen}
\usepackage{caption}
\usepackage{hyperref}
\usepackage{amsthm}
\usepackage{amstext}
\usepackage[total={6.5in,8.75in},top=0.75in, left=0.9in]{geometry}
\usepackage[utf8]{inputenc}
\usepackage{graphicx}
\usepackage{amsmath}
\usepackage{amssymb}
\DeclareMathOperator{\D}{\mathbf{D}}
\DeclareMathOperator{\df}{\mathbf{df}}

\newtheorem{definition}{Definition}[section]

\raggedright
\renewcommand{\section}[1]{%
\bigskip
\begin{center}
\begin{Large}
\normalfont\scshape #1
\medskip
\end{Large}
\end{center}}

\renewcommand{\subsection}[1]{%
\bigskip
\begin{center}
\begin{large}
\normalfont\itshape #1
\end{large}
\end{center}}

\renewcommand{\subsubsection}[1]{%
\vspace{2ex}
\noindent
\textit{#1.}---}

\renewcommand{\tableofcontents}{}

\bibpunct{(}{)}{;}{a}{}{,}  

\begin{document}
\begin{flushright}
Version dated: \today
\end{flushright}
\bigskip
\noindent Phylogenetic Derivative

\bigskip
\medskip
\begin{center}

\noindent{\Large \bf Phylogenetic Derivative: A Tool for Assessing Local Tree Reconstruction in the Presence of Recombination}
\bigskip



\noindent {\normalsize \sc Jacqueline Kane$^1$, Joseph Rusinko$^1$, and Katherine Thompson$^2$}\\
\noindent {\small \it 
$^1$Mathematics and Computer Science, Hobart and William Smith Colleges, Geneva, NY, 14456, USA;\\
$^2$Statistics, University of Kentucky, Lexington, KY, 40536, USA}\\
\end{center}
\medskip
\noindent{\bf Corresponding author:} Joseph Rusinko, Mathematics and Computer Science, Hobart and William Smith Colleges, 300 Lansing Hall,
Geneva, NY, 14456, USA; E-mail: rusinko@hws.edu.\\


\vspace{1in}

\subsubsection{Abstract} Recently, much attention has been given to understanding recombination events along a chromosome in a variety of field. For instance, many population genetics problems are limited by the inaccuracy of inferred evolutionary histories of chromosomes sampled randomly from a population. This evolutionary history differs among genomic locations as an artifact of recombination events along a chromosome. Thus, much recent attention has been focused on identifying these recombination points. However, many proposed methods either make simplifying, but unrealistic, assumptions about recombination along a chromosome, or are unable to scale to large genome-wide data like what has become commonplace in statistical genetics. Here, we introduce a \emph{phylogenetic derivative} to describe the relatedness of neighboring trees along a chromosome.  This phylogenetic derivative is a computationally efficient, flexible metric that can be also be used assess the prevalence of recombination across a chromosome. These proposed methods are tested and perform well in analyzing both simulated data and a real mouse data set. \\
\noindent (Keywords: recombination, phylogenetic trees, phylogenetic derivative, Robinson-Foulds distance, path interval distance )\\


\vspace{1.5in}
\section{Introduction and Background}
    Recent improvements in sequencing technology have resulted a plethora of genetic data available for analysis. However, these data have a specific, highly complex correlation structure \citep{rasmussen2014}.  For example, a recent study identified 71 genomic locations that were associated with risk of Crohn's disease.  However, in total these loci accounted for only 21.5\% of the estimated heritability of the disease \citep{zuk2012}. More specifically, ``genomic location,'' in this context, refers to a single nucleotide polymorphism (SNP), or a single DNA base pair that differs across observations under study. In another case, as of 2012, 52 SNPs had been associated with multiple sclerosis (51 of these associations post 2007) through analysis of genome-wide association study (GWAS) data, but these only attributed to 10\% of the genetic variation in MS \citep{visscher2012}. The sources of ``missing heritability'' in GWAS data have been widely debated in genetic analysis \citep{visscher2012}. One proposed explanation of this missing heritability is a lack of association mapping methods (those that aim to draw links between a genetic influence and trait) to identify small genetic effects.  Some recent association mapping methods aim to use the local evolutionary history at each SNP in order to inform inference of such associations \citep{zollner2005,mailund2006,besenbacher2009,pan2009,thompson2013,thompson2016,zhang2012}. Recently, phylogenetic methods have been proposed as one possible avenue of gaining more information out of existing SNP data sets \citep{mailund2006,besenbacher2009,thompson2013,roses2012}.

    Thus, improving phylogenetic tree estimation is crucial to gaining more information about SNP data. However, this is a non-trivial problem due to data set size (up to thousands of individuals and millions of SNPs). Further, estimation itself is difficult due to the presence of recombination points in sequence data. In the absence of such recombination, a single phylogenetic tree can describe the evolutionary history of all base pairs along the chromosome. However, in the case of recombination, a single binary tree can no longer represent the entire sequence along the chromosome. This, along with the inherent relationships both among individuals at a single base pair and among the evolutionary history of nearby base pairs, present statistical challenges for analyzing this data. Biologically, however, accurate tree estimation is crucial to understanding widespread population genetics topics \citep{rasmussen2014}. However, existing methods either are unable to build trees for thousands of individuals due to computational constraints, or are heuristic in nature. In this work, we propose a \emph{phylogenetic derivative} that uses tree distance measures to compare features of estimated trees that are imperative to understanding applications in association mapping and population genetics.

    In the absence of recombination, all SNPs along a chromosome share the same evolutionary history, and this history can be represented by a single phylogenetic tree.  However, if a recombination occurs, two chromosomes break and exchange a portion of their genetic sequence, meaning that neighboring trees on either side of this ``break point,'' or recombination point, will not share the same evolutionary history, and therefore, may differ in phylogenetic tree topology. Although ancestral recombination graphs (ARGs) can be used to represent all phylogenies along the chromosome in the presence of recombination, these are highly complex representations except for very small numbers of SNPs and observations \citep{zollner2005}.  Thus, some researchers choose to represent the evolutionary history of a set of SNPs as a corresponding set of phylogenetic trees at each SNP \citep{thompson2013}.
    
    Until recently, methods to estimate trees at particular SNPs, or local phylogenetic trees, have been limited in that they are either motivated by a biological or statistical model but do not scale well to large numbers of observations (e.g., TreeLD \citep{zollner2005}, MARGARITA \citep{minichiello2006}, or RENT \citep{wu2011}), or can handle large sample sizes but are fairly heuristic in nature (e.g., Blossoc \citep{mailund2006}). Recently, methods that are more statistically justified have been developed, including those based on the sequential Markov coalescent model (e.g., ARGweaver \citep{rasmussen2014}) and RENT+ \citep{mirzaei2016}. Here, the performance of Blossoc and RENT+ will be compared.
    
    Using local phylogenies as a motivation, accurate identification of recombination points is crucial in assessing performance of these tree estimation methods. In particular, since the evolutionary histories of SNPs on either side of a recombination point will be unrelated, and the evolutionary histories of neighboring SNPs between two recombination points will be more related, identifying recombination points would provide researchers with blocks, or set of neighboring SNPs, that share an evolutionary history.  These blocks could then be used to partition SNPs along a chromosome into sets of SNPs with similar evolutionary histories.  The identification of blocks of SNPs sharing a phylogeny would provide more focused information about the evolutionary history of the chromosome.
    
    To date, however, the study of identification of recombination points in SNP data has been limited to methods understanding recombination rates that are unable to analyze large genomic data sets \citep{arnheim2003,jeffreys2004,greenawalt2006,marttinen2011,martin2010} or methods that make unrealistic assumptions about recombination. For example, some methods do not consider recombination hotspots \citep{mcvean2004}, even though these have been described for at least a decade \citep{kauppi2004,jeffreys2005}. Recently, one method developed focused on identifying differences in recombination rates among admixed populations under study \citep{wegmann2011}, while another uses likelihood computations to estimate recombination rates \citep{auton2007}. Other methods focus on pedigree-based data, not the population-based samples as we consider here \citep{campbell2016,ferdosi2014}. In addition, we note that many references (e.g., in \citep{gusfield2010}) only have sufficient, but not necessary, conditions for identification of physical recombination locations.
    
    Rather than work to identify physical locations of recombination points, we aim to help researchers understand whether there are large or small amounts of recombination within a chromosome, or region of a chromosome, based on a data set. To address the limitations of the field thus far, we propose using a phylogenetic derivative to describe the changes of the tree structure along the chromosome. This will allow a more global view of analyses using local tree estimates. Second, the phylogenetic derivative has a strong association with the recombination rate, which can serve as a proxy estimate for the amount of recombination.  In this way, the use of changes of tree structure within neighboring SNPs along the chromosome can serve as a mechanism for analysis of tree estimation methods.
    
    Next, some key notation is described. Then, the proposed phylogenetic derivatives are defined and used to make comparisons of tree families. The methods, analysis, and results of the simulated data and real data studies are described in detail next. Lastly, a discussion of the implications of this work is provided.

\section{Mathematical Framework}

\subsection{Families of Trees Along a Chromosome}
Here, we introduce some notation for families of local phylogenetic trees as the foundation for the proposed phylogenetic derivative and applications. To begin, in general, a family of phylogenetic trees is defined as follows:
\begin{definition}
Let $\mathcal{F} = (T_1, T_2, \hdots, T_k)$ be an ordered family of trees. 
\end{definition}
\noindent Note that a family of trees may be used to represent phylogenetic trees along a chromosome. Each of these phylogenetic trees are associated with a nucleotide and show the evolution of that nucleotide among individuals. 

\begin{definition}
Suppose there is a family of trees such that there is a tree associated with every nucleotide along a chromosome. We define this set of trees to be the chromosomal family,
$$ \mathcal{C} = (T_1, T_2, \hdots, T_k),$$
where $k$ is the length of a chromosome in DNA base pairs.
\end{definition}

Next, consider a special case when no recombination occurred along a chromosome. Then there is a single, topologically identical tree for each nucleotide in the chromosomal family, $\mathcal{C}$. That is, $T_1 = T_2 = \hdots = T_k$. However, if any recombination occurs, we then have recombination sites, or recombination points, followed by regions of no recombination.
\begin{definition}
A non-recombining region is defined here as a set of adjacent nucleotides where no recombination has occurred. Thus, each nucleotide in this region is associated with a topologically identical tree. We define the family of trees representing a non-recombining region as $$\mathcal{R} = (R_1, R_2, \hdots, R_k).$$ 
\end{definition}

It then follows that with recombination present, $\mathcal{C} = (\mathcal{R}_1,\mathcal{R}_2, \hdots, \mathcal{R}_k)$. That is, the chromosomal family is made up of a set of non-recombining regions.
Next, we focus on SNPs along a chromosome. In particular, we define a SNP family of trees as follows.
\begin{definition}
We define a SNP family of trees to be $$\mathcal{S} = (S_1, S_2, \hdots S_k), $$ where each tree $S_1, \hdots, S_k$ is associated with a SNP.
\end{definition}
Note that not every non-recombining region may contain a SNP, and some non-recombining regions may contain multiple SNPs. 
\begin{definition}
We define an estimated SNP family of trees to be $$\mathcal{ES} = (ES_1, ES_2, \hdots ES_k), $$ where each tree $S_1, \hdots, ES_k$ is associated with a SNP.
\end{definition}
At times it is convenient to index the SNP and estimated SNP trees, not by the SNP number $(1,2,\cdots, k)$, but by the location of the associated SNP along the chromosome.  Unless otherwise noted we index using the former (SNP number) rather than the latter structure.

\subsection{Phylogenetic Derivatives}
In phylogenetics, the term tree distance or metric has two distinct meanings.  One notion refers to the distance between taxa, which is measured by some function of the DNA sequence data (Jukes Cantor for example).  Throughout this paper we refer to the second usage which is a comparison between phlyogenetic trees.  Two such examples include the Robinson-Foulds distance \protect\citep{robinson1981}, (which counts the number of splits which appear in one tree but not the other), and the path interval distance \protect\citep{huggins2012first,coons2016} (which  records the largest difference between the length of a path between a pair of taxa on one tree, with the length of the path between the same pair of taxa on the second tree). While this paper focuses on tree distances that only measure topological distances, all of the definitions proposed here also apply to tree distances that account for differences in branch-length as well (for example the path distance \protect\citep{steel1993distributions}).

First, consider the two families of trees $\mathcal{F} = (T_{1}, T_{2}, \hdots, T_{k})$ and $\mathcal{F}' = (T'_{1}, T'_{2}, \hdots, T'_{k})$. Note that $|\mathcal{F}| = |\mathcal{F}'| = k$. 
\begin{definition}
\label{def:avgtreedist}
 Given a tree metric $d$, we define the distance between families $\mathcal{F} = (T_1, T_2, \hdots, T_k)$ and $\mathcal{F}' = (T'_1, T'_2, \hdots, T'_k)$ as
$$d(\mathcal{F}, \mathcal{F}') = \frac{\sum_{i = 1}^{k}d(T_{i}, T'_{i})}{k}.$$
\end{definition}
This distance measure compares the two families $\mathcal{F}$ and $\mathcal{F}'$ tree-to-tree, and will only capture if the general trends are the same if trees are similar between the two families. Note that $d(\cdot)$ may refer to any distance measure throughout this discussion.

We introduce the \emph{phylogenetic derivative} as a tool for understanding the rate of change between trees in a family. Similar to how the derivative of a continuous function represents an instantaneous rate of change of that function, the phylogenetic derivative represents a discrete rate of change of a family of trees, or the change in tree topology between every pair of adjacent trees in $\mathcal{F}$.
\begin{definition}
For an ordered family of trees $\mathcal{F} = (T_1, T_2, \hdots, T_k)$ and a tree metric, $d$, we define the phylogenetic derivative as 
$$\D\mathcal{F} = (d(T_1,T_2), d(T_2,T_3), \hdots, d(T_{k-1}, T_k)).$$ We represent the elements of $\D\mathcal{F}$ as $(\df_1, \df_2, \hdots, \df_{k-1})$, or $\df_i = d(T_{i}, T_{i+1})$ for some $1 \leq i \leq k-1$.  Further we define the average derivative of $\mathcal{F}$, to be the average of the $\df_i$.
\end{definition}

This notion of change along the chromosome extends to higher order derivatives as well.
\begin{definition}
Let $\mathcal{F} = (T_1, T_2, \hdots, T_k)$ be an ordered family of trees and $\D\mathcal{F} = (\df_1, \df_2, \hdots, \df_{k-1})$ be the phylogenetic first derivative of $\mathcal{F}$ where $\df_i = d(T_i, T_{i+1})$ for some distance $d$ and $1 \leq i \leq k-1$. We define the phylogenetic higher order derivative as
$$\D^n\mathcal{F} = ((\df^{n-1}_2 - \df^{n-1}_1), (\df^{n-1}_3 - \df^{n-1}_2), \hdots, (\df^{n-1}_{k-1} - \df^{n-1}_{k-2})).$$ We denote the elements of $\D^n\mathcal{F}$ as $(\df^n_1, \df^n_2, \hdots, \df^n_{k-2})$. That is, $\df^n_i = (\df^n_{i + 1} - \df^n_i)$ for $1 \leq i \leq k-n$.  
\end{definition}
Note that $|\D^k\mathcal{F}| = k-n$, where $k = |\mathcal{F}|$. The phylogenetic higher order derivative allows us to more effectively compare the changes of the rates of change between families of trees, adding another dimension to the family comparisons that can be made using phylogenetic derivatives.

\begin{definition}
\label{def:avgderiv}
Let $\mathcal{F}$ and $\mathcal{F}'$ be ordered families of phylogenetic trees such that $|\mathcal{F}| = |\mathcal{F}'| = k$. We define the distance between phylogenetic derivative vectors $\D\mathcal{F}$ and $\D\mathcal{F'}$ as
$$d(\D\mathcal{F}, \D\mathcal{F}') = \frac{\sum_{i = 1}^{k-1} |\df_i - \df_i'|}{k-1} $$
\end{definition}

\noindent This distance measure determines how similar the rates of change between two families $\mathcal{F}$ and $\mathcal{F}'$ are. This can be used to gain information about amounts of recombination along the chromosome.

\subsection{Interpreting the derivative}
In the absence of recombination and if SNP-trees were estimated correctly, the phylogenetic derivative of a chromosomal family of trees would be uniformly zero.  Each recombination event should cause a change of the local tree topology at some SNP along the chromosome; thus, for the chromosomal family there would be a point where the phylogenetic derivative was non-zero. For neighboring trees in a SNP-family of trees, it is possible that multiple recombination events have occurred in between the SNP samples.  Therefore it is reasonable to interpret larger phylogenetic derivative values to indicate either a single, but very influential recombination event, or an accumulation of smaller recombination events between the sampled SNPs. Either way, one would anticipate a positive association between the amount of recombination between a pair of SNP trees, and the phylogenetic derivative value.

Next, we demonstrate the use of the phylogenetic derivative in three examples. These include examples in the case that the local phylogenetic trees at each SNP are known and the case in which the local phylogenetic trees at each SNP are estimated.

\subsubsection{Example 1: Chromosomal family with true phylogenetic trees}
As discussed above, a set of trees with no recombination yields a family of trees that are all topologically equivalent to one another. Thus, the derivative of this family contains all zero values. However, consider when recombination does occur. Suppose we have the chromosomal family 
$C = (\mathcal{R}_1, \mathcal{R}_2, \mathcal{R}_3, \mathcal{R}_4)$ made up of families trees from different non-recombining regions each of which contain exactly ten trees so that the total chromosome length is $40$ basepairs. It then follows that the phylogenetic derivative, $\D\mathcal{C}$, has exactly three non-zero entries $\df_{10},\df_{20},$ and $\df_{30}$.

\subsubsection{Example 2: SNP family with true phylogenetic trees}
Suppose as before that there are four distinct non recombining regions $C = (\mathcal{R}_1, \mathcal{R}_2, \mathcal{R}_3, \mathcal{R}_4)$ each of exactly $10$ base pairs.  Now assume we we have perfect SNP tree reconstruction, but only have access to SNPs from sites $5,15,18,35,36$ and $38$. Assume the local SNP-trees have been accurately reconstructed as $\mathcal{S} = (S_1, S_2, \hdots, S_6)$. Assuming perfect reconstruction we can also take the phylogenetic derivative of $\mathcal{S}$ which would be $(d(S_1,S_2),0,d(S_3,S_4),0,0).$ Here the zero derivative values are indicative that those pairs of SNPs occurred in the same non-recombining region and thus have the same corresponding true tree topology.   However, one might expect that $d(S_3,S_4)>d(S_1,S_2)$ as two recombination events have occurred between SNPs $3$ and $4$. The average derivative value would provide a general estimate of the amount of recombination.

\subsubsection{Example 3: SNP family with estimated phylogenetic trees}
Now consider the same SNP situation described in Example 2, but with a method of local SNP-tree reconstruction used to estimate the local trees. We still have $C = (\mathcal{R}_1, \mathcal{R}_2, \mathcal{R}_3, \mathcal{R}_4)$ and forty basepairs. Assume the local SNP-trees have been  reconstructed as $\mathcal{ES} = (ES_1, ES_2, \hdots, ES_6)$. We would then assume the estimated derivative value would be $(d(S_1,S_2) \pm \epsilon,0+\epsilon,d(S_3,S_4) \pm \epsilon,0 + \epsilon,0 + \epsilon)$.  Here the $\epsilon$ refers to the difference between the distance between the estimated trees, and the distance between the corresponding true trees.  Assuming that $\epsilon$ is small relative to the distances between the true trees, one should still anticipate detecting the recombination between SNPs $1$ and $2$, as well as that between SNPs $4$ and $5$. The average derivative values would still provide a general estimate of the amount of recombination. However this estimate is biased, since the epsilon values when the true trees match, can only be positive.

\section{Applications of Phylogenetic Derivatives}
Now that the behavior of the phylogenetic derivative has been illustrated, this section will show some applications of phylogenetic derivatives. These applications will be used to analyze both simulated and real data.

\subsection{Comparison of local phylogeny estimators}
In this section, we demonstrate how the phylogenetic derivative can be used to compare algorithms for local tree reconstruction. The simulations proposed here are intended to demonstrate the usefulness of the derivative as an analysis tool rather than to assess the quality of the reconstruction tools themselves.

In this work, we simulate a true evolutionary history for a chromosome, producing as an artifact of this, phylogenetic trees that are associated with each SNP along the chromosome. Thus, we have true SNP family of trees that is called $\mathcal{S}_T$. It then follows that these true trees can be compared with trees from programs that estimate SNP trees. We compare two programs, Blossoc and RENT+, that build SNP trees. Call the SNP family created from the Blossoc program $\mathcal{S}_B$, and call the SNP family created from the RENT+ program $\mathcal{S}_R$. By comparing the estimated SNP trees to the true SNP trees, we can determine the accuracy of the Blossoc and RENT+ as well as areas of strengths and weaknesses in the programs. However, note that RENT+ does not estimate the local phylogenetic tree for the first SNP along the chromosome, so there is no phylogenetic derivative for the first to second SNP comparison. This does not impact the overall results and comparisons made via the phylogenetic derivative.

In order to assess the performance of these distance metrics and tree estimation techniques, data were simulated to compare the performance of tree estimation techniques using the proposed methods. Specifically, data sets were simulated using the program, ms \citep{hudson2002}, which produces data for $n$ chromosomes from a population. Parameter settings were varied and included a mutation rate of $\mu=2.0 \times 10^{-10}$, recombination rates varying from $\nu=0$ (no recombination) to $\nu=10^{-8}$, for, $n=30$, $50$, $100$, or $250$ chromosomes (each of length 1,000,000 DNA base pairs) randomly sampled from a population of $N_0=20,000$ diploid individuals.  As in ms, we define the recombination rate parameter to be $4 N_0 \nu (\text{Chromosome Length} - 1 )$ in our simulations when the rate is varied. The resulting simulated data included phylogenetic trees at each DNA base pair and phased SNP data.

Next, local phylogenetic trees were estimated at each SNP using two methods: Blossoc \citep{mailund2006} and RENT+ \citep{mirzaei2016}. Resulting simulated data sets included the true local phylogenetic tree at each base pair location (and, by artifact, at each SNP), phased SNP data, as well as estimated local phylogenetic trees at each SNP from Blossoc and RENT+.

\begin{figure}[ht]
    \centering
    \includegraphics[width=0.90\textwidth]{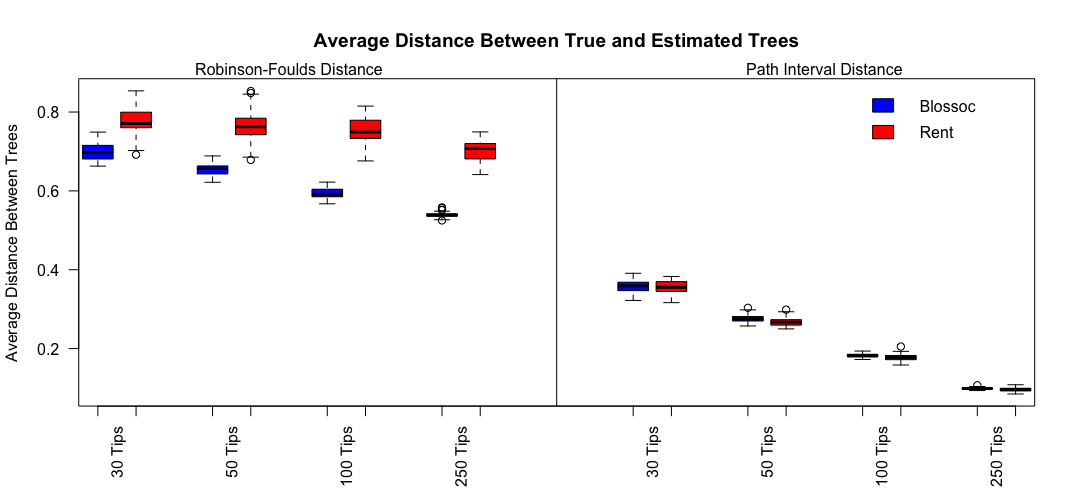}
    \caption{Average Distance Between True and Estimated Trees. This plot shows the average distance between true and estimated trees, as measured by each distance metric (left plot: Robinson-Foulds distance; right plot: path interval distance), for trees with varied numbers of tips. We note that with either distance metric, the distances between true and Blossoc trees (blue boxes) are smaller than the distances between the true and RENT+ trees (red boxes).  Results are based on simulated data sets.}
    \label{fig:avgdist1}
\end{figure}

For each simulated data set, let $\mathcal{S}_{T} = (T_1, T_2, \hdots, T_k)$ be the true SNP family of trees, $\mathcal{S}_B = (B_1, B_2, \hdots, B_k)$ be the Blossoc SNP family of trees, and $\mathcal{S}_R = (R_1, R_2, \hdots, R_k)$ be the RENT+ SNP family of trees. Note that $k$ varies across simulated data sets. Using the Robinson-Foulds distance metric, $RF$, we first calculated the distance, $d_{RF}(\mathcal{S}_T, \mathcal{S}_B)$ between the true SNP family and the Blossoc SNP family as well as the distance, $d_{RF}(\mathcal{S}_T, \mathcal{S}_R)$ between the true SNP family and the RENT+ SNP family. Similarly, using the path interval distance metric, $PI$, we calculated $d_{PI}(\mathcal{S}_T, \mathcal{S}_B)$ and $d_{PI}(\mathcal{S}_T, \mathcal{S}_R)$.  Figure \ref{fig:avgdist1} shows the average distance between trees, which is calculated using Definition \ref{def:avgtreedist}. Notice that using Robinson-Foulds distance, the average distance between true and Blossoc trees is smaller than the average distance between true and RENT+ trees. Using the path interval distance, both Blossoc and RENT+ have similar average tree distances from the true trees.

We then calculated the phylogenetic derivative using two tree distance metrics for each of the simulated data sets, $\mathcal{S}_T,$ $\mathcal{S}_B$, and $\mathcal{S}_R$. Again, we used the $RF$ distance to calculate the phylogenetic derivatives $\D_{RF}\mathcal{S}_T$, $\D_{RF}\mathcal{S}_B$, and $\D_{RF}\mathcal{S}_R$ and the path interval distance to compute the phylogenetic derivatives $\D_{PI}\mathcal{S}_T$, $\D_{PI}\mathcal{S}_B$, and $\D_{PI}\mathcal{S}_R$.

\begin{figure}[ht]
    \centering
    \includegraphics[width=0.90\textwidth]{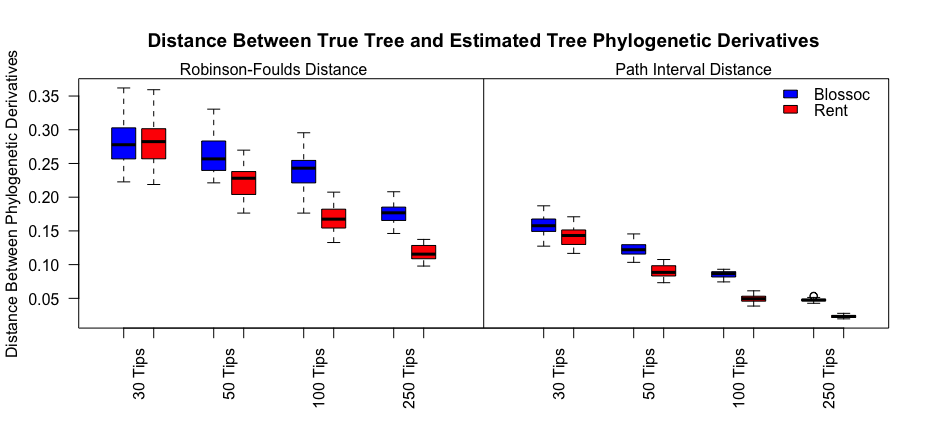}
    \caption{Distance between phylogenetic derivative vectors. This plot shows the average distance between phylogenetic derivatives for the true and estimated trees, as measured by each distance metric (left plot: Robinson-Foulds distance; right plot: path interval distance), for trees with varied numbers of tips. We note that with either distance metric, the differences between true and Blossoc-estimated trees (blue boxes) are larger than the differences between the true and RENT+-estimated trees (red boxes).  Results are based on simulated data sets.}
    \label{fig:peakvector1}
\end{figure}

We then use Definition \ref{def:avgderiv} to calculate the average distance between derivative vectors. Figure \ref{fig:peakvector1} shows the average distance between the Blossoc derivative vectors and the true derivative vectors as well as the average distance between RENT+ derivative vectors and the true derivative vectors using both Robinson-Foulds and path interval distance. Note that the error in the derivative estimation using RENT+ is smaller than the error when using Blossoc.

A more careful analysis of the data suggests that Blossoc's derivative values are uniformly smaller than those of RENT+. This appears to be an artifact of Blossoc's SNP-Tree reconstruction algorithm.  An analysis of individual runs shows that while the sizes of the peaks in Blossoc's derivative vector tend to be much smaller than those in both the true trees and in the RENT+ trees, the presence and absence of these peaks seems to mirror that of the true trees and RENT+ estimates.  This is phenomenon also occurs in the analysis of mouse chromosome data.

\subsection{Detecting recombination}
With perfect local phylogenetic tree reconstruction data, non-zero phylogenetic derivative values indicate the presence of recombination between observed SNPs.  In practice, where the local trees must be estimated, one would anticipate small phylogenetic derivative values in the non-recombining regions and large peaks, or sections of high derivative values in regions of the chromosome where there has been a large amount of recombination.

\begin{figure}[ht]
    \centering
    \includegraphics[width=0.90\textwidth]{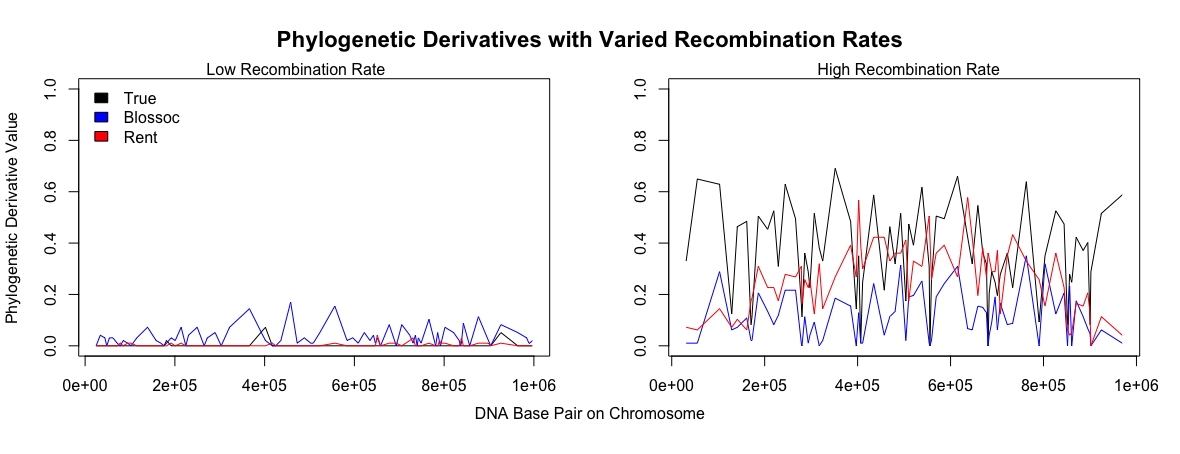}
    \caption{Comparison of phylogenetic derivatives with different recombination rates. The phylogenetic derivatives were calculated using the normalized Robinson-Foulds metric. The graph on the left shows three phylogenetic derivatives of SNP families with a recombination rate parameter of 1. The graph on the right shows three phylogenetic derivatives with a recombination rate parameter of 800. Red, blue, and black lines show the RENT+, Blossoc, and true trees, respectively. }
    \label{fig:low-highrecomb}
\end{figure}

We examine this phenomenon on test cases where the recombination rate is small, and thus the true phylogenetic derivative has relatively few non-zero entries, and when the recombination rate is high, where the true phylogenetic derivative has nearly all non-zero values. Figure \protect\ref{fig:low-highrecomb} shows that with a low recombination rate, the estimated Blossoc and RENT+ phylogenetic derivatives computed both identify some recombination events, but that noise in reconstruction makes the identification of recombination events a challenge for these tools. 

When the recombination rate is higher, as can also be seen in Figure \protect\ref{fig:low-highrecomb}, we see that both Blossoc and RENT+ phylogenentic derivatives identify the presence of a large amount of recombination. In addition, the derivative successfully follows the trend of where the SNP trees have larger or smaller topological differences although as in the case with small recombination rates, it is not yet possible to make precise readings identifying a specific subset of basepairs to be considered a recombination point.

 In addition to providing a visualization for recombination regions, the phylogenetic derivative can be used as a quantitative tool for measuring the recombination rate along a chromosome. In particular, the average value of the phylogenetic derivative can be used as a measure of the amount of recombination present along a region of the chromosome.  If one normalizes the tree distance, than the average phylogenetic derivative value ranges from $0$, indicating no-recombination across the region, to $1$ indicating so much recombination that each SNP has evolved completely independently from those around it.

\begin{figure}[ht]
    \centering
    \includegraphics[width=0.90\textwidth]{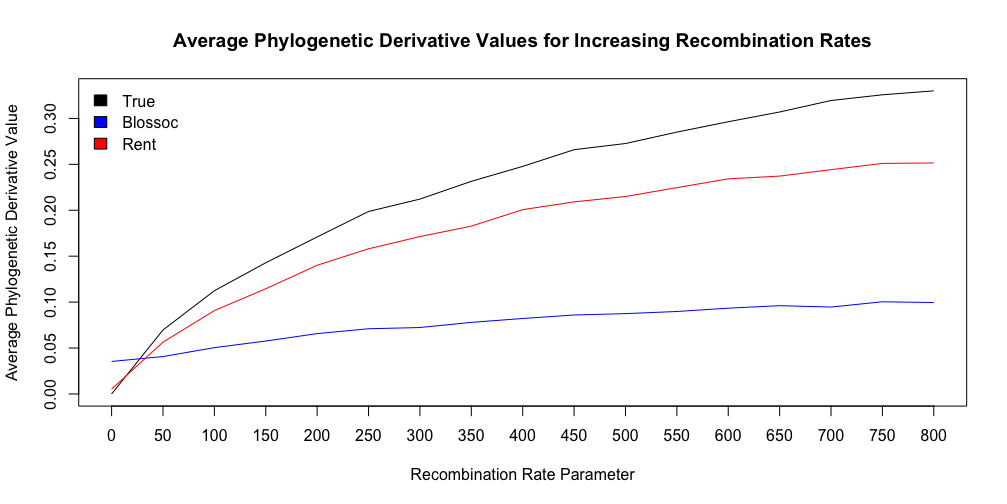}
    \caption{Average phylogenetic derivative values for increasing recombination rates. The phylogenetic derivatives were calculated using the Robinson-Foulds distance metric. Red, blue, and black lines show the RENT+, Blossoc, and true trees, respectively. Recombination rate parameter ranges from 0 to 800 in increments of 50, and for each recombination rate parameter value, 50 simulated data sets were analyzed.}
    \label{fig:incbarrier}
\end{figure}

Figure \protect\ref{fig:incbarrier} shows that when the number of taxa, and length of the chromosome are fixed, the average value of the phylogenetic derivative shows a clear positive association with the recombination rate itself. Thus, the average phylogenetic derivative might provide a mechanism for comparing and quantifying the amount of recombination along the chromosome.

\subsection{Real data analysis}

Real data analyzed here include SNP data from 288 male outbred mice from a 2012 study \citep{zhang2012b}.  Prior to analysis, these data were phased and missing data were imputed using Beagle version 3.3.2 \citep{browning2007}. The original goal of the study was to detect SNPs associated with various cardiovascular phenotypes, and this has been done using methods that estimate local phylogenies, as in \citep{thompson2013}. However, the goal of this work is to understand recombination across the chromosome, and to demonstrate the differences in behavior of tree estimation methods in the presence of few and many recombination events.  Thus, trees were estimated using Blossoc and RENT+.  These trees were analyzed using the distances described above.  Note that comparisons with true phylogenies are omitted since the true local phylogenies are unknown for real data.

\begin{figure}[ht]
    \centering
    \includegraphics[width=0.90\textwidth]{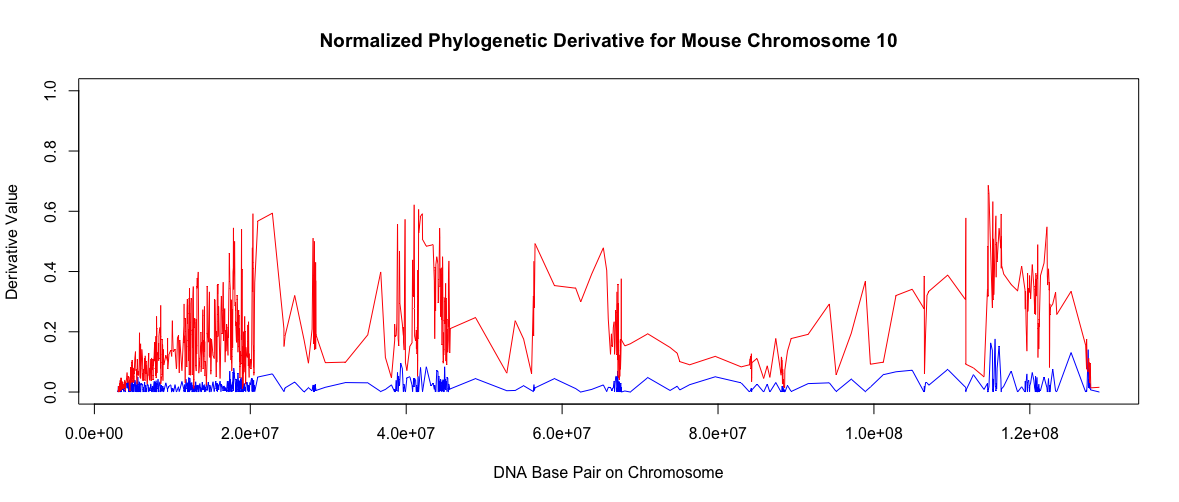}
    \caption{Phylogenetic derivative for trees from mouse chromosome 10. Both phylogenetic derivatives were calculated using Robinson-Foulds distance. Here, even when normalized, RENT+ (red line) shows larger values of the phylogenetic derivative than Blossoc (blue line) across the chromosome.}
    \label{fig:mousederiv10}
\end{figure}

\begin{figure}[ht]
    \centering
    \includegraphics[width=0.90\textwidth]{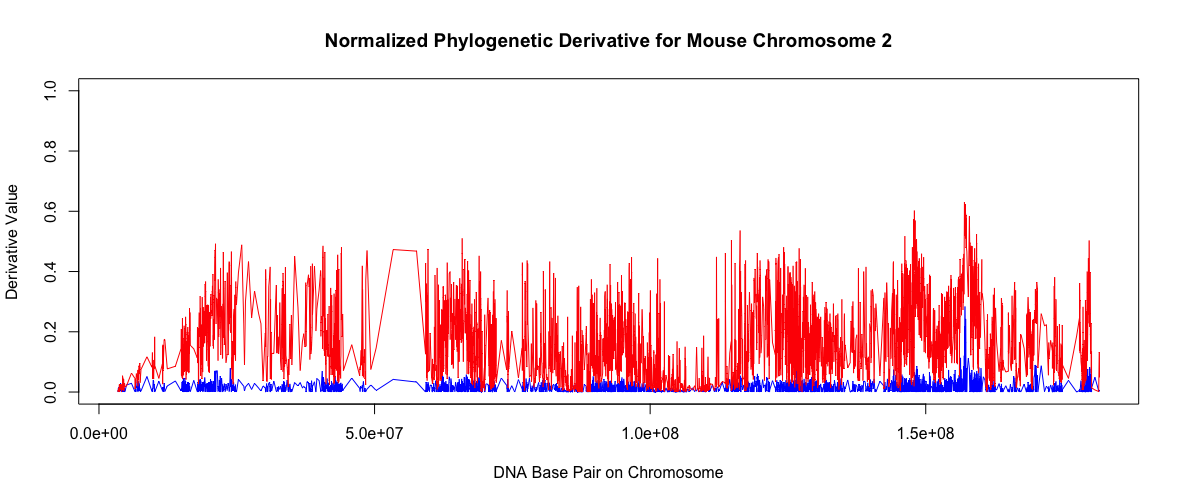}
    \caption{Phylogenetic derivative for trees from mouse chromosome 2. Both phylogenetic derivatives were calculated using Robinson-Foulds distance. Here, even when normalized, RENT+ (red line) shows larger values of the phylogenetic derivative than Blossoc (blue line) across the chromosome.}
    \label{fig:mousederiv2}
\end{figure}

Figure \ref{fig:mousederiv10} shows the phylogenetic derivative for the SNP family of estimated trees from mouse chromosome 10. In this data, chromosome 10 included information from 899 SNPs. Similarly, figure \ref{fig:mousederiv2} shows the phylogenetic derivative for the SNP family of trees from chromosome 2, which included information from 4609 SNPs. These two chromosomes were chosen due to their variation in size, so that data sets with large and small numbers of trees could be compared. The phylogenetic derivatives in Figures \ref{fig:mousederiv10}-\ref{fig:mousederiv2} show Blossoc trees having smaller phylogenetic derivative values with little variation. On the other hand, RENT+ trees have larger phylogenetic derivative values with more variation in the data. Notice however, that while the scale of the phylogenetic derivatives is different, both Blossoc and RENT+ indicate similar regions of recombination.

While an initial viewing of the phylogenetic derivative for chromosome 2 reads as if there is more recombination along this chromosome than in chromosome 10, some of this this perception is an artifact from chromosome 2 containing data from far more SNPs than chromosome 10, and thus opportunities for peaks in the derivative value.  We note, however that the average phylogenetic derivative for chromosome 10 is 18.46059 (0.01610871 normalized) for Blossoc and 216.8041 (0.1886893 normalized) for RENT+ in comparison to the average phylogenetic derivative value of 14.82726 (0.01293827 normalized) for Blossoc and 191.263 (0.1668962 normalized) for RENT+ for chromosome 2. Estimated trees from RENT+ may be indicating a larger amount of recombination than Blossoc along each of the chromosomes analyzed.   Both Blossoc and RENT+ indicate a higher average derivative value along chromosome $10$ than in chromosome $2$ indicating more recombination along chromosome $10$ a finding confirmed in \protect\citep{jensen2004comparative}.


\section{Discussion and Conclusion}

The phylogenetic derivative is a new tool for the analysis of local trees constructed across the chromosome. The derivative provides a framework for studying how SNP trees vary across the chromosome. In addition to its biological implications, the introduction of the phylogenetic derivative raises several mathematical questions, most interestingly how far can one extend the derivative analogy (optimization, integration, approximation) into the context of phylogenetic trees.

Based on the simulation and real data analyses presented here, the proposed metrics provide insight into the performance of local tree estimation algorithms. In particular, we see that the distance between phylogenetic derivatives can be used to assess how well local tree estimators detect changes in the trees along the chromosome.  Through this metric, we note that RENT+ tends to produce trees for which the phylogenetic derivatives are closer to those of the true local phylogenies than Blossoc does in a variety of simulation settings and using multiple tree-to-tree distance measures.  This is a very surprising finding, because using the same simulation data, Blossoc does a better job at reconstructing the local trees themselves.  Ideally, one would develop a local tree estimator which accurately reflects the trees themselves and the corresponding derivatives.  In the meantime we defer to individual practitioners in their choice of local tree estimation software, as our experiments do not justify the choice of Blossoc or Rent+ over the alternative.

Advantages of using the phylogenetic derivative include its flexibility to use different tree-to-tree distance measures.  For instance, our study focused on topological differences among trees, as these differences can be wildly detrimental to local phylogenetic tree estimation in association mapping. In one case, a comparison of analyses of true and estimated phylogenies showed that method performance was limited by the accuracy of the estimated phylogeny \citep{thompson2016}. However, should a user be interested in both topological and branch length differences among trees, another distance metric may be used in place of the Robinson-Foulds distance or Path Interval distance. 

Secondly, we see the phylogenetic derivative as a way to study recombination along the chromosome.  Using the fact that recombination points should produce different topological differences on either side of the break point, any such recombination event should yield non-zero phylogenetic derivative values. Our results demonstrate that both RENT+ and Blossoc can distinguish between data sets with high and low recombination rates.  However, these methods are not yet accurate enough to enable one to detect specific recombination events by analyzing the phylogenetic derivative.  Moreover, we observe a positive association between the average phylogenetic derivative in a family, and the underlying recombination rate in the data.  We suggest further study which may enable this relation to be more precisely quantified, which would enable practitioners to estimate the recombination rate in their data (or particular sub-regions of their data) directly from the estimated phylogenetic derivative.
    
Although this work addresses some of the limitations in identifying recombination events using SNP data in population-based samples, the potential to extend these methods to other tree estimation scenarios is inherent. For example, these methods may be extended to compare branch length estimates by using different distance measures between each pair of trees, or these methods may be useful in analyzing pedigree-based samples. In any case, the scores proposed here provide users with a versatile, computationally efficient method to explore recombination in genetic sequence data. More fully understanding aspects of recombination in SNP data may improve the estimation of local phylogenies at SNPs along a chromosome, which can begin to address the limitations of ARG inference in population genetics. Moreover, this improved understanding can contribute to addressing the challenges faced by limited knowledge of recombination behaviors in various types of sequence data.

 \section{Acknowledgements}
This material is based upon work supported by the National Science Foundation under Grant No. DMS-1358534. Further, the authors would like to thank Mark Curiel for helpful conversations in the initial phases of developing the phylogenetic derivative.

\bibliographystyle{sysbio}
\bibliography{refs}
\end{document}